\csname @addtoreset\endcsname{equation}{section}

 CUT HERE

\documentstyle[12pt]{article}

\input eqnsection.sty                   

\begin{document}

\centerline{\bf Anisotropies of the Cosmic Background}
\centerline{\bf Radiation in a Reionized Universe}
\vskip 19pt
\baselineskip=15pt
\centerline{Robin Tuluie\footnote{current address: Department of Astronomy and
Astrophysics and Center for Gravitational Physics and Geometry, The
Pennsylvania State University, University Park, PA 16801}
 \& Richard A. Matzner }
\centerline{\it Department of Physics and Center for Relativity}
\centerline{\it The University of Texas at Austin, Texas 78712}
\centerline{ }
\centerline{Peter Anninos}
\centerline{\it National Center for Supercomputing Applications}
\centerline{\it Beckman Institute}
\centerline{\it 405 N. Mathews Ave.}
\centerline{\it Urbana, IL 61801}
\centerline{ }

\vskip 19pt
\centerline{\bf Abstract}
\vskip 9pt

{\small
We trace the evolution of cosmic microwave background photons propagating
through a model universe filled with `hot' dark matter and a reionized
intergalactic medium (IGM).
 The reionization of the intergalactic medium is achieved by
UV photons emitted from the decaying `hot' dark matter neutrinos.
The model universe is represented by a comoving cube of present
 size $529 Mpc$ through
which we propagate photons and subject them to the Sachs-Wolfe effect and
Thomson scattering off the ionized fraction of the IGM,
beginning at $z=1900$ just before the epoch of decoupling until the present
$z=0$. The simulation follows the evolving matter inhomogeneities and
microwave background photons into the present nonlinear regime and
yields  temperature maps of the
microwave sky as a comoving observer would detect. Our temperature maps
display anisotropies on scales ranging from 2 to 8 degrees with rms amplitudes
ranging from $\Delta T/T_{rms} =2.8$ to $3.4 \times 10^{-5}$ for
$\Omega_{\circ}=1$ and $h=0.55$   without
reionization.
If we allow the neutrinos to decay and subsequently
reionize the IGM, the level of the temperature anisotropies
is reduced to $\Delta T/T_{rms} =0.8$ to $1.1 \times 10^{-5}$.
This includes the Doppler shifts suffered by the
 CMB photons as they scatter off the
moving ionized fraction of the IGM. If we turn off the contribution to
$\Delta T/T_{rms}$ from these Doppler shifts,  the temperature anisotropies
are reduced  to
$\Delta T/T_{rms} =0.7$ to $0.8  \times 10^{-5}$  for $\Omega_{\circ}=1$ and
$h=0.55$ in the reionized model.
If, however, reionization is caused by a late and sudden energy input into the
 IGM, we find that some reduction in $\Delta T/T$ occurs,
 but not enough to make this scenario
consistent with observations.
Including the contribution to $\Delta T/T_{rms}$ from Doppler shifts at the
surface of last scattering, the level of anisotropies in this scenario is
usually at the same level and sometimes even higher than the original
primordial fluctuations.
We find that reionization via the decaying neutrino hypothesis is more
 efficient
at reducing primordial anisotropies and generates smaller Doppler shift
anisotropies than a model in which the universe is reionized via a late
and sudden energy input into the IGM.
In the case the universe is not reionized, we find that the contribution to
$\Delta T/T_{rms}$ from Doppler shifts at decoupling is small compared to the
Sachs-Wolfe effect at the large angular scales we are considering.}

\vskip 49pt
\section{Introduction}
\vskip 19pt

Recent measurements of the cosmic microwave background radiation (CMBR)
impose severe constraints on current matter evolution scenarios.
  One would expect the
CMBR to carry a signature of the evolving large scale  matter structures
of the early universe due to the gravitational redshifting of the CMBR
photons through the Sachs-Wolfe effect (Sachs \& Wolfe 1967).
  However, no temperature
fluctuations above a level of $\Delta T/T_{rms}=3.5\times10^{-5}$
 at $\leq 1^{\circ}$  angular scales have been observed yet (Meinhold
\& Lubin 1990, Timbie \& Wilkinson 1990, Meinhold et al. 1991,
Devlin et al. 1992),
while on large angular scales ($\geq 7^{\circ}$)
 COBE found $\Delta T/T_{rms} = 1.1
\times 10^{-5}$ (Smoot et al. 1992).
In the standard `cold' dark matter (CDM) evolutionary
scenarios, the large scale distribution of matter would not reach the
level suggested by recent measurements (Broadhurst et al. 1990,
Maddox et al.1990)
without violating these CMBR constraints. A `hot' dark matter (HDM) model
is better suited to generate large scale structures constistent with
 observations, but lacks small scale power unless additional seeds
 (such as cosmic strings) for galaxy formation  or a non-gaussian spectrum
 of initial perturbations is invoked. It is also severly constrained by
the COBE data, which requires $\Omega_{\circ} h^2 \sim 1$ in this model,
 where
$\Omega_{\circ}$
is the density parameter and $h$ is the value of the Hubble constant
 in units
of $100 km/s/Mpc$
(Anninos et al. 1991): for choices of
parameters  $\Omega_{\circ} h^2 < 1 $ the HDM model exceeds COBE
limits unless an efficient mechanism of erasing the temperature fluctuations,
such as a reionization of the intergalactic medium, has  taken
place.

In this paper we investigate a `hot' dark
matter model with $\Omega_{\circ}=1.0$ and $h=0.55$ in which a
reionization of the intergalactic medium (IGM) occurs. The simulation
model used to evolve the dark matter is similar to our previous
treatment (Anninos et al. 1991). Our model universe
consists of a succession of periodically joined, three dimensional cubes,
of length 529 Mpc resolved with $64^3$
cells.
We use a Cloud-in-Cell particle-mesh model (Hockney \& Eastwood 1981)
for the dark matter simulation
in which physical quantities such as potential, density and force are
defined on a uniformly spaced cubical grid.

  The mass in our model consists of `hot'
dark matter such as neutrinos, plus a small baryonic component which
traces the dark matter. We assume a metric of the form
\begin{eqnarray}
ds^2 &=& -(1+2\phi)dt^2+a(t)^2(1-2\phi)(dx^2+dy^2+dz^2),
\end{eqnarray}
where $a(t)$
is the homogeneous background expansion factor and $\phi \ll 1$
is the Newtonian gravitational potential.
Matter is therefore treated as in
perturbation theory, and we neglect terms of second
order ($\phi^2$) and smaller. In this Newtonian limit, the equations
of motion for the matter  are  solved
along with Poisson's equation for the gravitational potential.

 The initial inhomogeneity of this collisionless
matter takes the Harrison - Zel'dovich spectral shape
 (Zel'dovich 1970) modulated by
Landau damping that produces a short wavelength cutoff.
 The short wavelength cutoff associated with the HDM spectrum is achieved
 via the neutrino transfer
function $T(k)$ (Doroshkevich \& Khlopov 1981, Centrella et al. 1988, Anninos
et al. 1991). This models the Landau damping of perturbations below a
scale corresponding  to the horizon
size at the epoch of matter-radiation equilibrium
\begin{eqnarray}
\lambda_c&=&13(s \Omega_{\circ} h^2)^{-1} Mpc,
\end{eqnarray}
where $s$ is chosen as a length scale that defines the number
of Mpc per grid zone. As $\Omega_{\circ} $ and $h$ are varied, the comoving
size of perturbations changes as $(\Omega_{\circ} h^2)^{-1}$.
 In order to obtain
a similar resolution within the cube, different values of $\Omega_{\circ} h^2$
require different size cubes as it is desirable to preserve the total
number of perturbation wavelengths in the cube when comparing to our previous
simulation with $h=1$.

The initial particle velocities
are set equal to the growing mode velocities
found from Zel'dovich's linear solution (Zel'dovich 1971).
The amplitude of the initial fluctuations
is chosen such that the two-point correlation function
\begin{eqnarray}
\xi(r) &=& \left(\frac{r}{r_0}\right)^{\gamma}
\end{eqnarray}
reaches the observed value for  clusters
of $\gamma \sim -1.8$ at a redshift of $z=0$
for separations $r \sim 25 h^{-1} Mpc$ (Bahcall \& Soneira 1983).
 Note that this is the same
value for $\gamma$ as that observed for galaxies with separations
$r \le 15 h^{-1}$ (Peebles 1980).
 Execution of the simulation is terminated at this point.
The spatial scale is then determined from the constraint that we preserve
the total number of perturbations in the cube, which yields $s=8.26 Mpc/zone$
and a computational domain of $529 Mpc$.
We have also verified that the final matter
 velocity fields in the cube (at z=0)
are consistent with the observed large scale velocity fields (Tuluie 1993a).
 This gives
us an additional check on the validity of our termination of the simulation.
The final structure resembles
 the sheet/filament/knot
structure observed in the universe, including large voids of matter, as well
as large scale peculiar and bulk velocities
 (Anninos et al. 1991, Centrella et al. 1988).

\vskip 19pt

Beginning at an initial redshift of $z=1900$ (just prior to the
beginning of the recombination epoch), we allow a
 representative sample of photons
to propagate through this developing gravitational structure.  Rather
than dealing with an ensemble of waves, photons are propagated
individually.  We subject the individual photons to Thomson scattering
at times chosen in a probabilistic way and at a rate determined by the
instantaneous local mean free path in the model.
  The instantaneous local mean free path is determined by the
 {\em local} baryonic density and not the
background value. The scattering
process, done in the rest frame of the scattering electron, is
accomplished via a Monte Carlo simulation using the Thomson cross section.
We do not compromise our model by choosing an instantaneous epoch of
recombination, but rather trace each photon continuously through the
epoch of recombination, preempting any ambiguity in
the choice of surface of last scattering.  We assume that electrons and
protons trace the dark matter by defining the baryonic density for each
zone as a fraction $\Omega_b$ of the instantaneous local dark
matter density for that particular zone.  For the purpose of computing
electron densities for photon mean free paths we assume that the
fraction of the density parameter due to baryons, $\Omega_b$
ranges from 0.02 to 0.2.

In addition to the gravitational
redshifting of the CMBR  and the Thomson scattering occuring during decoupling
we now include the effects of repeated Thomson
scatterings of the
radiation off the IGM  in the case when the universe undergoes an epoch of
reionization of the IGM between the decoupling and present eras
 (sections 2 \& 3).
We also include the Doppler shifts suffered by the CMBR  as
it scatters off the moving IGM  both during the decoupling and during the
 reionization
eras of the universe (section 4).

\vskip 19pt

It is generally thought that reionization of the IGM at late times could
smooth temperature variations of the CMBR in models which would
otherwise predict temperature anisotropies above current observational
levels.  However, even in an extreme scenario in which a late and sudden
maximal
reionization occurs, we find from our computations
that  such a model for reionization is not able   substantially to
 reduce temperature anisotropies in the CMBR unless the reionization
begins at an unphysically early epoch $(z \geq 60)$, a time at which
the existence of discrete ionizing sources is unlikely (Giroux 1993).
A fully ionized IGM at this early stage
could also interfere with the onset of star formation (Vishniac 1993).
In this manner we are
able to rule out a low Hubble constant hot dark matter model of the
universe even if a late and sudden reionization of the IGM should occur.

However, should reionization of the IGM be accomplished through the decay of
the HDM, e.g. decay of the $29eV$ neutrinos as in our model,
 CMBR temperature anisotropies
could  be reduced sufficiently to make this particular realization of the HDM
 model viable again (Scott et al. 1991).
 In particular, for a density parameter of $\Omega_{\circ}=1$ and
Hubble constant of $H_{\circ}=55km/s/Mpc$ we
find that the standard, nonreionized
 model exceeds COBE temperature measurements of the CMBR.
By allowing the neutrinos to decay, the resulting UV radiation produces a low
(a few percent at large redshifts), but gradually increasing degree of
ionization in the IGM. We find that this `early and gradual' scheme of
reionization is effective in reducing CMBR temperature anisotropies on scales
up to {\em twice} the particle horizon scale
{\em at the surface of last scattering}
(Weinberg 1972, Ellis \& Rothman 1992). This scale can be substantially larger
 in a reionized universe than the comparable scale at
recombination, which is $\theta_*\simeq 2z_{ls}^{-1/2} \simeq 3^{\circ}
 \mbox{ for
large z, with }z_{ls}=z_{rec}=1500 $ (eq.(2.4) of this paper).
 For example, the angular
 size of the particle horizon at the surface of last scattering in the
$h=.55$, $\Omega_b=0.2$ universe
reionized
by neutrino decay is $\simeq 5^{\circ}$, so $\theta_*\simeq 10^{\circ}$
 at $z_{ls} \simeq 160$ (eq. (2.3) and
fig.1). Because the surface of last scattering is rather thick at
late times, i.e. $\delta z_{ls} \sim z_{ls}$, the above result is only an
approximate estimate.

The angular range over which reionization is effective in erasing primordial
 temperature fluctuations is in fact somewhat smaller than this
estimate due to the random walk
 of the photons through the surface of last scattering, which tends to shorten
the effective horizon size. Our simulation takes this into account as we
 follow the individual photons along their
 scattering histories. We are able to give
precise values for the effectiveness of various scenarios of reionization in
 reducing primordial temperature fluctuations in the CMBR. We present our
results in the form of sky maps of the CMBR such as a comoving observer might
 make, ranging in size from $2^{\circ} \times 2^{\circ}$ to $8^{\circ}
 \times 8^{\circ}$. Simulations
on scales larger than $8^{\circ}$ are currently not sensible, because
 the statistics for
the photon maps become poor and because
the matter structures in the comoving cube are
imaged repeatedly.

We find that the `early and gradual' (EG) scheme of
reionization based on the UV radiation from the decaying neutrinos is far more
effective in reducing CMBR temperature anisotropies
 than the `late and sudden' (LS) scenario. Besides being less hypothetical
 about the source of the ionizing radiation and its effect on the IGM  than is
the LS  scenario, the EG
scheme makes contact with a variety of cosmological
 observations  and is thus ultimately (and perhaps very soon) falsifiable
through independent observations (Sciama 1990a-d).
We note in addition that tests of reionization, such as the Vishniac effect
(Vishniac 1987),
may be less stringent in universes which undergo an early and gradual
reionization history and could thus possess anisotropies below current
detector limits.

\vskip 49pt
\section{Reionization via Decaying Neutrino}
\vskip 19pt

The previously described simulation model (Anninos et al. 1991),
in particular the simulation of
CMBR anisotropies, lends itself as an ideal testbed for modeling various
scenarios of reionization.

Not only are we able to model and image the Sachs-Wolfe effect, but we can also
include the temperature distortions due to Doppler shifts suffered by the CMBR
photons from  the
peculiar motions of the scattering electrons,
 and we also take into account the finite thickness
 of the surface of last scattering. During an epoch of
reionization of the IGM all of the above effects are important in erasing
the primordial temperature patterns  over
 a wide range of angular scales while at the same time
generating secondary anisotropies (Kaiser 1984, Peebles 1987,
 Efstathiou 1988, Vishniac 1987).
 In a future paper we will describe
and estimate the effect of non-linear contributions to $\Delta T/T$ (Vishniac
1987) specific to our particular model (Tuluie et al. 1993).

Our goal is not to fine tune reionization to yield a particular realization
of the HDM model that is consistent with observations (As already noted,
the generic HDM model
has trouble generating the observed small scale structures unless there are
special seeds for galaxy formation), but rather  to
provide a testbed in which various scenarios of reionization can be evaluated
accurately for their effectiveness in reducing CMBR anisotropies without
generating excessive second order fluctuations.

\vskip 19pt

The first model for reionization considered here is the
`Early and Gradual' (EG) model. It relies on the UV radiation
from the decaying dark matter as its source for the ionizing radiation.
In this model  a finite lifetime of the dark matter neutrino
  allows a small fraction of the HDM to  decay into UV photons,
 which subsequently reionize the IGM.
 Depending on whether the decay $ \nu \longrightarrow \gamma
\; + \; \nu_1 $ yields a massive or massless neutrino $\nu_1$, the resulting
photon will have energies of up to $m_{\nu}/2$. For this photon to ionize
hydrogen, the  energy $E_{\gamma}$
 of the photon must be larger than the threshold ionization energy
 required for hydrogen in the ground state,
 so $m_{\nu} > 27.2eV$, but below bounds set
by astrophysics.

A more stringent lower bound on $m_{\nu}$ can be set by
 requiring that the decay photons from decays occuring in the Coma cluster
arrive at our galaxy with energy exceeding $13.6eV$ and thus are absorbed
heavily
 by interstellar neutral hydrogen, in which case the the null result of UV
photons from dark matter decay (Holberg et al. 1985) is satisfied. The lower
limit on $m_{\nu}$ is therefore $E_{\gamma} > 13.6 eV \cdot (1+z_{Coma}) =
13.9 eV$, or $m_{\nu} > 27.8 eV$ (Sciama 1990b).

 The most constraining upper bound comes from the limit of the
 metagalactic ionizing flux from an intergalactic cloud in Leo (Reynolds et al.
 1986, Songaila et al. 1989), which requires $E_{\gamma} \leq 15 eV$ and, if
$m_{\nu} \gg m_{\nu_1}$, yields $ m_{\nu} \leq 30 eV$.
The combined bounds thus constrain $m_{\nu}=28.9 \pm 1.1eV$
(Sciama 1990b). In addition, the decay time of the neutrino is $t_d=
y\cdot10^{23}sec$,
with a prefered value of $y=1.5$ (Sciama 1990a).

For our simulation we choose a neutrino mass of $29eV$ and decay time of $1.5
\cdot 10^{23}sec$, which is  within the astrophysical bounds.
The effect that the decaying dark matter has on the matter evolution in our
simulation is negligible, since only a fraction $(\sim 10^{-6})$  of neutrinos
 has decayed
by the present time.

\vskip 19pt

Next we consider the contribution to the ionized fraction of the IGM,
$X_{e,\nu}$,
that is due to the ionizing radiation from the dark matter decay.
Ionizing photons will shortly after their emission be absorbed by the neutral
fraction of hydrogen of the IGM (which is $\sim 90$ \% after recombination).
The decay rate of  the dark matter, $n_d t_d^{-1}$, thus balances the
recombination
rate $X_{e,\nu} n_b t^{-1}_{rec}$,
where $t_{rec}=5.4 \cdot 10^{16} X_{e,\nu}^{-1} \Omega_b^{-1} (1+z)^{-5/2}$sec,
 of the IGM (Scott et al. 1991):
\begin{eqnarray}
X_{e,\nu}&=&\frac{n_d}{n_b}t_d^{-1}t_{rec} \nonumber\\
&=&55.7y^{-1/2}\left(\frac{\Omega_b}{.1}\right)^{-1}(1+z)^{-5/4},
\end{eqnarray}
where $n_d$ and $n_b$ are the dark matter and baryonic matter densities and
$y$ is the decay time parameter.
$X_{e,\nu}$ follows this result until it reaches unity at a redshift
\begin{eqnarray}
z_i&=&24y^{-2/5}\left(\frac{\Omega_b}{.1}\right)^{-4/5}
\end{eqnarray}
In the standard recombination scenario (Peebles 1967) the ionization
fraction $X_e$ is very small for $z > $ few hundred ($X_e \sim 10^{-4}$).
Hence the interaction
of $X_{e,\nu}$ with $X_e$ is negligible since $X_{e,\nu}$ itself
doesn't grow appreciably until $z < $ few hundred.
The total ionization fraction $X_{e,t}$  can therefore
 be written as the sum of the contributions from the standard input
 $X_e$ and the contribution from the
decaying neutrino input $X_{e,\nu}$. A more accurate
result for $X_{e,t}$ requires a numerical integration of
$dX_{e,t}/dt = dX_e/dt + dX_{e,\nu}/dt$
 (Salati \& Wallet 1984, Asselin et al. 1988),
 but we have verified that this  yields
little improvement for our range of values for $\Omega_b,
\Omega_{\circ} $ and $h$.

The results for $X_{e,t}$ for various values of $\Omega_b$ are
 plotted in fig.2. Comparing to Asselin et al. (1988) we see
 that this is nearly identical to the result for a numerical integration.
 Fig.3 shows the optical depth in our model as a function of
redshift. Rather than using an approximate analytical result, the optical depth
in our model is computed from the instantaneous value of the baryonic density
integrated along the photons' history (including
variations in $\Omega_b$ due to the changing neutrino densities resulting from
structure formation). In this way, scattering occurs more frequently in
densely clustered regions of our model, which tends to further contribute to
the reduction of the Sachs-Wolfe effect generated from these regions.

We consider baryonic densities of $\Omega_b =0.2$ and $\Omega_b =0.02$. A lower
value of $\Omega_b$ tends to result in a higher degree of ionization, hence
the optical depth decreases somewhat as $\Omega_b$ decreases, but not as fast.
We find the surface of last scattering (i.e. the point at which the optical
 depth reaches unity) located  at a redshift of
$z_{ls}=160$ for $\Omega_b =0.2$ and
 $z_{ls}=380$ for $\Omega_b =0.02$. The critical scale on which reionization is
effective in erasing primordial temperature fluctuations is roughly
twice the
particle horizon scale at the surface of last scattering.
This statement is qualitative in view of the finite thickness of
the surface of last scattering.
 The proper distance of this horizon is (G.F.R. Ellis \& T.
Rothman 1992, Weinberg 1972)
\begin{eqnarray}
d_H(t_0)&=&R(t_0)\int_{0}^{t_0} \frac{c`,dt}{R(t)} =
2cH_0^{-1}(1+z_{ls})^{-3/2} \nonumber\\
&=&6000h^{-1}(1+z_{ls})^{-3/2}Mpc \nonumber
\end{eqnarray}
during the $\Omega_{\circ}=1$ matter dominated era.
The critical angular scale below which reionization is effective is  twice the
angular scale of this proper distance:
\begin{eqnarray}
\theta_*&=&2\theta_H(t_0) =  2 \arcsin \left(\frac{\sqrt{1+z_{ls}}}{z_{ls}+
 (1- \sqrt{1+z_{ls}})} \right) \\
&\simeq&2z_{ls}^{-1/2} \qquad \mbox{for $z \gg 1$}.
\end{eqnarray}

For CMBR photons the effective damping scale is actually somewhat smaller than
this due to the random walk of the photons through the surface of last
scattering, which tends to shorten the horizon size.
Our simulation takes this into account as we follow
the individual photons along their scattering histories. Any decreasing effect
reionization suffers on larger angular scales is thus modeled accordingly.

This angular damping  scale  $\theta_*$ is plotted in fig.1 as a function of
redshift. Included in
this plot is the location of the surface of last scattering. We see that
the damping  scale at the surface of last scattering is
$\theta_* \sim 3^{\circ}$ for standard recombination with no reionization.
If reionization takes place, $\theta_*$ can be as large as
$\sim 10^{\circ}$ (for $z_{ls}=160$), which is
sufficiently large to affect the COBE $7^{\circ}$ results.

\vskip 19pt

The CMBR temperature maps are displayed in plates 1-8 and the results of these
and other simulations are listed in tables 1-7.
Temperature maps are constructed by projecting the redshifts suffered
by the photons onto a plate with a $60 \times 60$
grid; there are thus $\sim 280$ photons per grid for $10^6$ photons.
We have used a 32 level color scale.
The color in each zone of the map is
determined by averaging the redshifts ($\Delta T / T$) of the
photons which fall in the zone, so that each zone has a
characteristic redshift. Each plate is filtered by a
smoothing algorithm
which averages the nearest neighbors on the sky and thus provides
a smoother map. Features larger than one cell are
stable under this smoothing.
The color images are displayed with a resolution of
$343 \times 343$ pixels by interpolating the $60 \times 60$
smoothed grid data. Photon redshifts are represented as red colors ,
while photon blueshifts appear in blue on the color maps.

Table 1 shows the results of a simulation of the CMBR sky as
viewed from an observer located in
the center of our computational cube,
looking into a randomly selected
direction of our model universe with a
$4^{\circ} \times 4^{\circ}$ window size. We find
temperature anisotropies of
 $(\Delta T/T)_{rms} =  2.8 \times 10^{-5}$ for
 $\Omega_b =0.02$ and $(\Delta T/T)_{rms} = 3.1 \times 10^{-5}$ for
$\Omega_b =0.2$.
 if no reionization takes place.
 These temperature anisotropies are a  factor $\sim 3$ above COBE results and
thus do not present a viable scenario. Simulations in other directions
and various angular scales give similar results and are listed in table 1.

If reionization by the decaying neutrinos does occur,
the  fluctuations for the same parameters are reduced
to  $(\Delta T/T)_{rms} = 1.0 \times 10^{-5}$ for $\Omega_b =0.02$ and
$(\Delta T/T)_{rms} = 0.8 \times 10^{-5}$ for $\Omega_b =0.2$ (table 2).
In table 2 a new column, $N_p$, the number of photons used in the simulation,
appears. The values corresponding to $N_p=\infty$ are the physically predicted
ones that correspond to the value for $\Delta T/T_{rms}$ if the data for
our finite number of photons is extrapolated to an infinitely large ensemble
(see also section 4).

Note that the above result includes the contribution to $\Delta T/T$ arising
from Doppler shifts incurred by the CMBR photons from the peculiar motions
of the electrons at the surface of last scattering.
This contribution is small in
the standard scenario, but becomes noticeable during an epoch of reionization,
which has a considerably more recently situated surface of last scattering.
This is discussed in section 4.

 If we turn off the
 Doppler shifts suffered by the CMBR as it scatters off the ionized fraction
of the IGM (see section 4) in order to ascertain the effect of reionization
alone, the level of the anisotropies is reduced to
$(\Delta T/T)_{rms} = 0.8 \times 10^{-5}$ for $\Omega_b =0.02$ and
$(\Delta T/T)_{rms} = 0.7\times 10^{-5}$ for $\Omega_b =0.2$
on a $4^\circ$ scale (table 3).
This shows that in a reionized scenario the CMBR anisotropies
are due  both to  the Sachs-Wolfe effect and to
the Doppler shifts at these angular scales.
 The dependence of $(\Delta T/T)_{rms}$ on $\Omega_b$
is mostly due to the
different locations of the surface of last scattering and the associated
values for the optical depth (fig.3).
 {\em We find that reionizing the IGM by the decaying neutrinos reduces
anisotropies by at least a factor $\sim 2$ even if the Doppler shifts
 the CMB photons suffer
 during reionization are included.}

 The temperature maps corresponding to the reionization scenarios (plates 3-8)
 show a general smoothing of the primordial
structures,  with the underlying  pattern of these structures
 (as visible in the standard, non-reionized
 scenarios of the corresponding plates 1 and 2) reduced but
 still discernible. This is especially noticable when comparing between
plates 1 and 3 as well as between plates 2 and 4.
The contribution to $(\Delta T/T)_{rms}$ from the Doppler shifts is
listed in table 6.
It is apparent that Doppler shifts generate additional structure in the
 CMBR maps  on sub $1^{\circ}$ scales when comparing plate 6, which
shows the gravitational temperature shifts only,
to plate 7, which shows Doppler shifts only. The angular
size of the bulk velocity fields at the surface of last scattering
in our simulation is at roughly the same scale.

The effect reionization has on the CMBR should begin to taper off for angles
larger than twice the horizon at the surface of last scattering, which is
$\theta_*  \sim 6^{\circ}$ for
$\Omega_b =0.02$ and $\theta_*  \sim 10^{\circ}$ for
$\Omega_b =0.2$ in this model. However, currently our simulations
are limited to  $ \sim 8^{\circ}$ or smaller
scales as both the statistics and small scale resolution of the photon maps
suffer at larger angles and we notice
 repeated structure in our photon
maps at large angles reflecting the finite size
 of our simulation domain. In the future we
will run a $128^3$ version of this model, with a correspondingly larger
computational domain and the ability to explore super $\sim 10^{\circ}$ scales.

In addition, on sub $1^{\circ}$ scales nonlinear
 contributions to $\Delta T/T$ arising from
Doppler shifts of the CMBR photons scattering
off the ionized fraction of the IGM as it falls into the DM potential wells
become important as the IGM reionizes (Vishniac 1987). Although we cannot
resolve this effect currently, we could estimate its signature in the
CMBR sky based
on our knowledge of $v_{pec}$, $\delta \rho/\rho$ and $X_e$ at the surface of
last scattering in our model (Tuluie et al. 1993b).

\vskip 49pt
\section{`Late and Sudden' Reionization}
\vskip 19pt

This particular scheme of reionization assumes that at some epoch $z_{start}>5$
the IGM became fully ionized rather quickly, say within one expansion time.
Prior to $z_{start}$, the ionization fraction was at its low,
post-recombination
value of $X_e \sim 10^{-4}$.
 Even though sources, such as quasars, primeval galaxies, or
population III stars for such a
scenario have been postulated, their existence
prior to $z \sim 60$ is doubtful (Shapiro 1986a,b, Shapiro \& Giroux 1987).
In addition, it appears that at least
some of these
sources would be incapable of providing the required degree
of ionization at early epochs.
In particular, if the remaining density in the IGM is sufficently large,
there are not enough known
quasars to have ionized the IGM. In fact, for $\Omega_b=0.1$,
they fail by an order of magnitude (Giroux 1993).

Our purpose is to show that for reasonable scenarios for the starting redshift
of reionization in a late and sudden
reionization model of the IGM, i.e. $z_{start} \leq 60$,
 primordial CMBR anisotropies are not reduced
sufficiently to make our hot dark matter models compatible with COBE and other
microwave observations. In fact, for our choices of $z_{start}$ and $\Omega_b$,
 this scheme (LS) proves less effective than the EG scenario discussed
 earlier.
In addition, LS scenarios can carry a larger signature of
secondary distortions  due to the higher degree of ionization
at the surface of last scattering.

\vskip 19pt

The evolution of the ionization fraction in the LS scenario is shown in fig.4,
along with the optical depth. At the epoch $z_{start}$ the value for $X_e$
jumps to unity as the photoionizing sources turn on. We have neglected
evolutionary effects  which might delay this process in lieu of considering
a maximal scenario where $X_e$ evolves as pictured; this assures a maximal
reduction of $\Delta T/T$ in this model.

We have allowed $X_e$  to increase to unity  independent
of the local density peaks which usually spawn the discrete sources of
  photoionizing radiation. By doing so we hope to ascertain the maximum
 effect a global
reionization of the IGM can have in reducing the anisotropies of the
CMBR. It also avoids the biasing ambiguity associated with discrete sources
of reionization.

Our results are presented in tables 4 and 5 and plate 8.
 Table 4 lists values for $\Delta T/T$
corresponding to the sky maps.
As before, we have included the contribution to the anisotropy in $\Delta T/T$
resulting from Doppler shifts of the CMBR scattering off the moving IGM.
 Table 5 is as table 4 except that Doppler shifts
have been turned off.

In general, the reduction in $\Delta T/T$ from the standard, non-reionized
scenario (table 1) is weaker than what we had seen in the EG scenario (table
2).
In particular, at best a slight  reduction is evident for
 most runs, which is not sufficient to meet current CMBR anisotropy
measurements
(Smoot et al. 1992).
This is less than what was obtained in the EG scenario.
{\em Reionization by a late and sudden photoionizing
 epoch thus is less effective
at reducing primordial CMBR anisotropies than a scenario in which the HDM is
allowed to decay and subsequently reionize the IGM.} In fact, due to the
contribution to $\Delta T/T$ from the late Doppler shifts, the LS
scenario can in some instances generate secondary anisotropies that are at
the same level as the
originally present primordial fluctuations.

A more realistic model for the LS scenario
 would take into account evolutionary effects
as well as the discrete nature of the ionizing sources through
appropriate biasing.
 We expect such a model to be even less successful
at reducing primordial temperature anisotropies
 than the maximal one considered here.

\vskip 49pt
\section{Temperature Anisotropies from Doppler Shifts}
\vskip 19pt

At any time during the evolution of the CMB photons through the forming
matter structures the photons can Thomson scatter
 off the free electron fraction of the IGM.
 In the standard, non-reionized scenarios the probability for
scattering after decoupling ($z \leq 1000$) is low,
 while in the reionizing scenarios
this probability can be substantial. At late times, the forming matter
structures can exhibit
substantial bulk motions relative to the comoving background, which can induce
Doppler shifts on the CMB photons scattered off this moving medium. Typically,
the Doppler shifts suffered at the surface of last scattering dominate the
contribution from previous Doppler shifts due to the larger peculiar velocities
at late times as well as the higher correlation of the velocity field.
The Doppler shifts suffered at the surface of last
scattering imprint a signature in the CMBR at angular scales
below the horizon scale at the surface of last scattering.

For the HDM model the contribution to $\Delta T/T_{rms}$ will
be small in the standard
scenario due to the smaller bulk velocities at the surface of last
scattering (i.e. at decoupling) and the Sachs-Wolfe effect
dominates the CMBR anisotropies in this model.
However, should a reionization of the IGM occur in this model (as discussed in
sections 2 and 3 of this paper), the induced
fluctuations on sub-horizon scales can be a substantial fraction of the
primordial anisotropies.
The contribution to $\Delta T/T$ resulting from Doppler shifts that occur
during the scattering process is calculated next.

The
scattering process is performed in the rest frame of the
scattering electron whose velocity is given by the local weighted
average of the velocities of matter overlapping the zone in which
the photon is scattered. Photons in comoving coordinates
are Lorentz transformed into the rest frame of the scattering
electron via the usual formula
\begin{eqnarray}
e^j &=& \frac{\beta\gamma n^j+(\gamma-1)n^i n^j {e^i}'+
\delta^{ij} {e^i}'}
    {\gamma(1+\beta n^i {e^i}')},
\end{eqnarray}
where  $e^i$ are the
photon directional cosines, primes refer to the
corresponding comoving quantities and $n^i=
v^i/\left|{\bf v}\right|, \beta = \left|{\bf v}\right|/c,
$and $\gamma=1/\sqrt{1-\beta^2}$ all refer to the electron.

The scattering process is accomplished with Monte-Carlo probabilistic
techniques. The scattering angles ($\Theta,\Phi$) are defined as
spherical coordinates relative to the incident ray in the electron
scatterer's rest frame. We choose $\Theta$ from a distribution in
$\cos \Theta$ that
varies as $\sim (1+\cos^2{\Theta})$. The
azimuthal angle $\Phi$ is chosen from a uniform distribution
over $0 \le \Phi \le 2\pi$.
We refer the reader to Anninos et
al. (1991). During the scattering process the
frequency of the photon does not change as observed in the electron frame.
However, after the scattering is complete in the electron's frame, the
photon is  Lorentz transformed back into the background frame.
 As the photon direction before and after the scattering
process in the electron frame differs in general, the transformation
back into the background frame results in a net frequency shift.

We find this frequency shift from our knowledge of the photon directional
cosines $e_I^i$ in the proper rest frame of the
total matter (i.e. the background frame)
before scattering, the photon directional
cosines $e_F^i$ in the
electron frame after scattering and the proper electron frame velocity
$\beta^i=
v^i/c$. The first transformation, from the background frame to the
 electron frame, yields a Doppler shift
\begin{eqnarray}
\frac{\omega'}{\omega}&=&\gamma - \gamma \beta_i e_I^i.
\end{eqnarray}
After the photon scatters, the transformation from the electron frame back into
the background frame yields an additional shift of
\begin{eqnarray}
\frac{\omega''}{\omega}&=&\gamma \frac{\omega'}{\omega}+
\gamma \beta_je_F^j \frac{\omega'}{\omega}.
\end{eqnarray}
Substituting for $\omega'/\omega$ and keeping terms to first order in $\beta_i$
yields
\begin{eqnarray}
\frac{\omega''}{\omega}&=&1-\beta_i(e_I^i-e_F^i).
\end{eqnarray}
This is the Doppler shift suffered by the photon in proper coordinates,
with proper
electron velocities $\beta^i$ and photon directional cosines
$e^i_I$ and $e^i_F$.

This Doppler shift is then included in the total redshift, which is
measured in proper coordinates
\begin{eqnarray}
1+z&=&\frac{a_r}{a_e}\left[1+(l+\phi)\mid^e_r \right]\frac{\omega}{\omega''},
\end{eqnarray}
where $l$ is the perturbation to the photon energy from the Sachs-Wolfe effect
(Anninos et al. 1991),
 $\phi$ is the gravitational potential redshift
  and the last factor is the
contribution to the redshift resulting from the Doppler shift.
The total fractional change in the microwave background temperature due to
all these effects is
\begin{eqnarray}
\frac{\Delta T}{T}&=& \frac{\delta z}{1+z} \\
&=&\phi_e-\phi_r-\int_r^e
\frac{2 {\bf e} \cdot {\bf \nabla} \phi}{a} dt +
{\bf \beta} \cdot ({\bf e}_I-{\bf e}_F),
\end{eqnarray}
where $\phi_e-\phi_r$ is the gravitational
 potential difference between emission and reception,
 the integral is the Sachs-Wolfe effect along the photon's
history  and the last term is  the
Doppler shift.

We model the photon evolution by propagating a large number
(typically $N_p=10^6$) of individual photons
though our model cube. We have found
 it neccessary to increase the number of photons when investigating
scenarios that consider recently reionized cases with Doppler shifts
 up to $N_p=8 \times 10^6$. This is neccessary to minimize the error
 arising from randomly sampling regions with substantial
electron velocities. If the electron velocities are as
 small as they are during recombination, then even a
 few samples in a given direction average to a value
which becomes negligible compared to the Sachs-Wolfe effect.
 With late reionization, the sampled Doppler shifts are larger
 by a factor $(1+z)^{1/2}$.
 We have verified that this effect indeed scales as
$1/\sqrt{N_p}$ for values of $N_p$ ranging from $10^5$ to
$8 \times 10^6$ and converges to the values presented
 in the tables (under $N_p= \infty$). Fig. 5
shows this convergence.

To include the Doppler shifts the CMB photons suffer during
 an epoch of reionization, in addition to the
gravitational temperature shifts suffered through the
Sachs-Wolfe effect, at the last timestep that the photon scatters we compute
the
Doppler shift from eq. (4.7)
  and add this to the previous value for $\Delta T/T$.
We therefore include all the Doppler shifts
 occuring at the surface of last scattering (SoLS),
 taking into account both the finite thickness of this surface
and the varying bulk velocity fields throughout the simulation domain.
We neglect previous Doppler shifts because
  Doppler shifts suffered at the SoLS dominate previous Doppler shifts
due to the higher peculiar velocities  and higher correlation of the
velocity field at the SoLS.

We have also considered the effects of the Doppler shifts alone on the CMBR,
turning off the Sachs-Wolfe effect for this case. The results are
 listed in table 6.
We see that no apparent visual correlation
 exists between CMBR anisotropies caused
by Doppler shifts  and CMBR anisotropies caused by the Sachs-Wolfe effect.
We can  compute the expected uncorrelated value of $\Delta T/T$ from
\begin{eqnarray}
\frac{\Delta T}{T}_{uncorr}&=&\sqrt{\left(\frac{\Delta T}{T}\right)
^2_{Dp} +\left(\frac{\Delta T}{T}\right)^2_{Gr}}
\end{eqnarray}
where $(\Delta T/T)_{Dp}$
 is the contribution
from the Doppler shifts only
and $(\Delta T/T)_{Gr}$ includes purely gravitational temperature distortions
only. Table 7 compares our values  and shows good argeement with eq. (4.8).
This is to be expected since during the scattering process any directional
correlation is destroyed by the randomly chosen value for the azimuthal
scattering angle.

\vskip 49pt
\section{Conclusion}
\vskip 19pt

We have traced the evolution of cosmic microwave
 background photons propagating
through a HDM model universe
with a reionized
intergalactic medium. Throughout the evolution,
 beginning at $z=1900$ until
$z=0$, we have subjected the CMB photons to gravitational potential
redshifts, to the Sachs-Wolfe effect, to Thomson scattering off the
ionized fraction of the IGM and to the Doppler
 shifts incurred during this
scattering from the bulk motions of the forming matter structures.
 We have found that for parameters of $\Omega_{\circ}=1$ and $h=0.55$
 the CMBR temperature anisotropies are above current
 observational constraints if no reionization occurs.

We have investigated two possible scenarios of reionization. We found
that a model which
suffers an early and gradual (EG) reionization of the IGM, as caused
 by the photoionizing UV radiation emitted by the decaying HDM, is sufficiently
efficient at reducing primordial anisotropies to the level observed by COBE. In
fact, our model skymaps agree well with COBE results.
This was achieved without the introduction of  any free parameters, as all
parameters in the model are determined by astrophysical observations
independent from CMBR observational constraints: the
 initial amplitude of the density perturbation spectrum is determined
 by the current slope of the matter correlation function (Anninos et.
al. 1991, Centrella et al. 1988), while
the free parameters for the neutrino mass and lifetime are constrained by
astrophysical bounds (Sciama 1990a-d). We have also shown
 that the effective scale at which reionization can dampen
 primordial anisotropies can be substantially
larger than the horizon size at decoupling. A HDM model reionized via a
decaying neutrino thus presents a viable alternative to other large scale
scenarios.

The second scenario of reionization investigated was a `best case' model for a
late and sudden (LS) scenario of reionization.
 In this model, the ionization fraction
of the IGM increases from $X_e \sim 10^{-4}$ to unity suddenly. This
late and sudden scenario proved unable to reduce primordial CMBR
temperature anisotropies sufficiently to meet current observations unless
the reionization occured before a redshift of $z \sim 60$. It is unlikely that
distinct photoionizing sources existed prior to this epoch in sufficient
numbers to create a fully ionized IGM (Giroux 1993). We conclude that this
particular model, even in our `best case' scenario, is less effective at
reducing primordial CMBR
temperature anisotropies than the EG model and does not meet current CMBR
temperature anisotropy observations.

\vskip 19pt

In addition, we have investigated the effect of the late, nonlinear matter
 condensation's signature in the microwave sky, models in which $\Omega_{\circ}
 \neq 1$, as well as the importance
of general relativistic
corrections to our Newtonian limit, and we
will present these results in a future
paper.

\vskip 19pt

{\em Acknowledgements} The authors especially thank
Ethan Vishniac
 for helpful discussions and a critical reading of the manuscript. R. T.
thanks D. W. Sciama for helpful discussions.
This research was supported by NSF grants PHY-8806567 and PHY-9310083 and
 Texas Advanced Research Program grant 085 and by a Cray University
Reasearch Grant to Richard Matzner. Computations reported
here were carried out on a Cray-2 and a Convex C3880 machine
at the National Center for Supercomputing
 Applications (University of Illinois).

\newpage

The notation in the tables is as follows:
$N_p=\infty$ corresponds to the value for $\Delta T/T_{rms}$ if the data for
our
finite number of photons is
extrapolated to an infinitely large ensemble.
Scenarios that do not have a value
for $N_p=\infty$ listed already have adequate
 resolution and the extrapolated value is very close to the listed one.
The minimum and maximum values for $\Delta T/T$ quoted are with respect
to a zero average temperature for each plate.

The plate numbers of the temperature color maps correspond to the plate numbers
in the tables, which provide information
 regarding the parameters for each plate.
 Plates 1, 2, 5, 6, 7 and 8 have the extreme red and blue colors of
 the color map independently set equal to the maximum
 value for the redshift and maximum value for the blueshift respectively.
Plates 3 and 4 have the extreme red and blue colors of the
 color map set equal to the maximum
 value for the redshift and maximum value for the blueshift of plates 1 and 2
respectively.
This
 way the reduction in the CMBR anisotropies due to reionization is
 directly visible when comparing plate 1 to plate 3 and  plate 2 to plate 4.

\newpage
\centerline{\bf Table 1}
\vskip 1pt
\centerline{Temperature Fluctuations in the Standard Scenario $\times 10^5$}
\centerline{($N_p=10^6$, $\Omega_{\circ}=1.0$, $h=0.55$, no reionization)}
\vskip 5pt
\centering
\begin{tabular}{ccccccc}   \hline\hline
{\em $\Omega_b$} & {\em Direction} & {\em Angle} & $\Delta T/T_{rms}$ & $\Delta
T/T_{min}$ & $\Delta T/T_{max}$ & {\em Plate} \\ \hline
0.02 & Random & $4^o$ & 2.8 & -6.4  & 7.6  & 1  \\
0.2  & Random & $4^o$ & 3.1 & -6.6  & 7.4  &    \\
0.2  & Random & $2^o$ & 2.8 & -7.2  & 7.4  &    \\
0.2  & Random & $8^o$ & 3.4 & -8.9  & 10.9 & 2  \\
0.02 & Corner & $4^o$ & 3.3 & -10.4 & 7.5  &    \\ \hline\hline
\end{tabular}
\vskip 49pt
\centerline{\bf Table 2}
\vskip 1pt
\centerline{Temperature Fluctuations in the `EG' Scenario
 $\times 10^5$, Incl. Doppler Shifts}
\centerline{($\Omega_{\circ}=1.0$, $h=0.55$, Doppler Shifts included in $\Delta
T/T$)}\vskip 5pt
\centering
\begin{tabular}{cccccccc}   \hline\hline
{\em $\Omega_b$} & {\em Direction} & {\em Angle} & $N_p$ & $\Delta T/T_{rms}$
 & $\Delta T/T_{min}$ & $\Delta T/T_{max}$ & {\em Plate} \\ \hline
0.02 & Random & $4^o$ & $ 10^5        $ & 2.6 & -7.8  & 8.7 &    \\
0.02 & Random & $4^o$ & $5 \times 10^5$ & 1.6 & -4.4  & 5.4 &    \\
0.02 & Random & $4^o$ & $ 10^6        $ & 1.5 & -4.5  & 5.0 &    \\
0.02 & Random & $4^o$ & $2 \times 10^6$ & 1.3 & -3.5  & 4.6 &3,5 \\
0.02 & Random & $4^o$ & $ \infty      $ & 1.0 &       &     &    \\
0.02 & Random & $8^o$ & $2 \times 10^6$ & 1.4 & -4.3  & 4.5 & 4  \\
0.02 & Random & $8^o$ & $ \infty      $ & 1.1 &       &     &    \\
0.02 & Random & $2^o$ & $2 \times 10^6$ & 1.1 & -3.4  & 3.6 &    \\
0.02 & Random & $2^o$ & $ \infty      $ & 0.9 &       &     &    \\
0.2  & Random & $4^o$ & $2 \times 10^6$ & 1.4 & -4.8  & 5.8 &    \\
0.2  & Random & $4^o$ & $  \infty     $ & 0.8 &       &     &    \\
0.02 & Corner & $4^o$ & $2 \times 10^6$ & 1.3 & -3.6  & 3.2 &    \\
0.02 & Corner & $4^o$ & $  \infty     $ & 0.9 &       &     &    \\
 \hline\hline
\end{tabular}
\newpage
\centerline{\bf Table 3}
\vskip 1pt
\centerline{Temperature Fluctuations in the `EG' Scenario
$\times 10^5$, no Doppler Shifts}
\centerline{($\Omega_{\circ}=1.0$, $h=0.55$, `EG' reionization)}
\vskip 5pt
\centering
\begin{tabular}{cccccccc}   \hline\hline
{\em $\Omega_b$} & {\em Direction} & {\em Angle} & $N_p$ & $\Delta T/T_{rms}$
& $\Delta T/T_{min}$ & $\Delta T/T_{max}$ & {\em Plate} \\ \hline
0.02 & Random & $4^o$ & $ 10^5        $ & 0.9 & -2.6  & 2.8 &    \\
0.02 & Random & $4^o$ & $2 \times 10^6$ & 0.8 & -2.2  & 2.1 &  6 \\
0.2  & Random & $4^o$ & $ 10^5        $ & 0.8 & -2.2  & 2.7 &    \\
0.2  & Random & $4^o$ & $2 \times 10^6$ & 0.7 & -2.0  & 2.1 &    \\
0.2  & Random & $8^o$ & $ 10^5        $ & 0.8 & -2.3  & 3.3 &    \\
0.2  & Random & $2^o$ & $ 10^5        $ & 0.8 & -2.8  & 3.1 &    \\
0.2  & Corner & $4^o$ & $ 10^5        $ & 0.7 & -2.2  & 1.9 &    \\
\hline\hline
\end{tabular}
\vskip 49pt
\centerline{\bf Table 4}
\vskip 1pt
\centerline{Temperature Fluctuations in the `LS' Scenario
 $\times 10^5$, Incl. Doppler Shifts}
\centerline{($\Omega_{\circ}=1.0$, $h=0.55$,
 `LS' reionization, Doppler shifts included in $\Delta T/T$)}
\vskip 5pt
\centering
\begin{tabular}{ccccccccc}   \hline\hline
{\em $\Omega_b$} & {\em Direction} & {\em Angle} & $N_p$ & $z_{rh}$ &
 $\Delta T/T_{rms}$ & $\Delta T/T_{min}$ & $\Delta T/T_{max}$ &
 {\em Plate} \\ \hline
0.02 & Random & $4^o$ & $2 \times 10^6$ & 60  & 6.0 & -14.6 & 17.2&    \\
0.02 & Random & $4^o$ & $ \infty      $ & 60  & 5.9 &       &     &    \\
0.2  & Random & $4^o$ & $2 \times 10^6$ & 60  & 2.5 & -5.7  & 8.2 &  8 \\
0.2  & Random & $4^o$ & $ \infty      $ & 60  & 1.9 &       &     &    \\
\hline\hline
\end{tabular}
\vskip 49pt
\centerline{\bf Table 5}
\vskip 1pt
\centerline{Temperature Fluctuations in the `LS' Scenario
 $\times 10^5$, no Doppler Shifts}
\centerline{($\Omega_{\circ}=1.0$, $h=0.55$, `LS' reionization)}
\vskip 5pt
\centering
\begin{tabular}{ccccccccc}   \hline\hline
{\em $\Omega_b$} & {\em Direction} & {\em Angle} & $N_p$ & $z_{rh}$ &
$\Delta T/T_{rms}$ & $\Delta T/T_{min}$ & $\Delta T/T_{max}$ &
{\em Plate} \\ \hline
0.02 & Random & $4^o$ & $10^5$ & 60  & 2.3 & -6.6  & 6.7 &    \\
0.2  & Random & $4^o$ & $10^5$ & 60  & 0.7 & -2.6  & 2.1 &    \\ \hline\hline
\end{tabular}
\newpage
\centerline{\bf Table 6}
\vskip 1pt
\centerline{Temperature Fluctuations in Various Scenarios
 $\times 10^5$, Doppler Shifts Only}
\centerline{($N_p=2 \times 10^6$, $\Omega_{\circ}=1.0$, $h=0.55$,
 only Doppler Shifts included in $\Delta T/T$)}
\vskip 5pt
\centering
\begin{tabular}{ccccccccc}   \hline\hline
{\em $\Omega_b$} & {\em Direction} & {\em Angle} & $N_p$ &
{\em Scenario} & $\Delta T/T_{rms}$ & $\Delta T/T_{min}$ &
 $\Delta T/T_{max}$ & {\em Plate} \\ \hline
0.02 & Random & $4^o$ & $1 \times 10^6$ &  `EG'   & 1.2 & -3.4 & 4.6  &    \\
0.02 & Random & $4^o$ & $2 \times 10^6$ &  `EG'   & 1.1 & -2.9 & 4.0  &    \\
0.02 & Random & $4^o$ & $4 \times 10^6$ &  `EG'   & 1.1 & -2.6 & 4.3  &    \\
0.02 & Random & $4^o$ & $8 \times 10^6$ &  `EG'   & 1.0 & -2.5 & 3.9  &  7 \\
0.02 & Random & $4^o$ & $   \infty    $ &  `EG'   & 0.9 &      &      &    \\
0.2  & Random & $4^o$ & $2 \times 10^6$ &  `EG'   & 1.2 & -4.2 & 4.3  &    \\
0.02 & Random & $4^o$ & $2 \times 10^6$ & `LS60'  & 5.6 & -13.0 & 15.3&   \\
0.2  & Random & $4^o$ & $2 \times 10^6$ & `LS60'  & 2.5 & -6.6  & 7.8 &   \\
 \hline\hline\end{tabular}
\vskip 49pt
\centerline{\bf Table 7}
\vskip 1pt
\centerline{Correlation Between Gravitational
 Redshifts and Doppler Shifts $\times 10^5$}
($N_p=2 \times 10^6$, $\Omega_{\circ}=1.0$, $h=0.55$,
 refering to eq. (4.8) and above tables)
\vskip 5pt
\centering
\begin{tabular}{ccccccccc}   \hline\hline
{\em $\Omega_b$} & {\em Direction} & {\em Angle} & {\em Scenario}
 & $\Delta T/T_{Gr}$ & $\Delta T/T_{Dp}$ & $\Delta T/T_{uncorr}$
 & $\Delta T/T_{rms}$  \\ \hline
0.02 & Random & $4^o$ & `EG'   & 0.8 & 1.1 & 1.4 & 1.3 \\
0.2  & Random & $4^o$ & `EG'   & 0.7 & 1.2 & 1.4 & 1.4 \\
0.02 & Random & $4^o$ & `LS60' & 2.3 & 5.6 & 6.1 & 6.0 \\
0.2  & Random & $4^o$ & `LS60' & 0.7 & 2.5 & 2.6 & 2.5 \\ \hline\hline
\end{tabular}

\newpage
\centerline{\bf References  }
\vskip 29pt

\halign{#&\quad# \hfil \cr

&Anninos, P., Matzner, R. A., Tuluie, R. and Centrella, J. 1991, \cr
&{\it Ap. J.} {\bf 382}, 71 \cr
\noalign{\vskip10pt}

&Asselin, X. et al. 1988, {\it Nucl. Phys.} {\bf B310}, 669 \cr
\noalign{\vskip10pt}

&Bahcall, N. A. and Soneira, R. M., 1983, {\it Ap. J.} {\bf 270}, 20 \cr
\noalign{\vskip10pt}

&Broadhurst, T., Ellis, R., Koo, D. and Szalay, A. 1990, \cr
&Nature  {\bf 343}, 726 \cr
\noalign{\vskip10pt}

&Centrella, J., Gallagher, J., Melott, A., and Bushouse, H. 1988, \cr
&{\it Ap. J.} {\bf 333}, 24 \cr
\noalign{\vskip10pt}

&Devlin, M. et al. 1992, Proceedings of the NAS colloquium on physical
cosmology
\cr
\noalign{\vskip10pt}

&Doroshkevich, A. G., and Khlopov, M. Yu., 1981, {\it Soviet. Astr.} {\bf 25},
  521 \cr
\noalign{\vskip10pt}

&Efstathiou, G. 1988, In {\it Large Scale Motions in the Universe} \cr
&e.d. V. C. Rubin and G. V. Coyne, 115 \cr
\noalign{\vskip10pt}

&Ellis, G. F. R. and Rothman, T. 1993, {\it Lost Horizons}, \cr
&University of Cape Town Preprint 1992/12 \cr
\noalign{\vskip10pt}

&Giroux, M. 1993, PhD. Thesis, University of Texas \cr
\noalign{\vskip10pt}

&Hockney, R., W., and Eastwood, J. 1981, {\it Computer Simulation Using} \cr
&{\it Particles} (New York: McGraw-Hill) \cr
\noalign{\vskip10pt}

&Kaiser, N. 1984, {\it Ap. J.} {\bf 282}, 374 \cr
\noalign{\vskip10pt}

&Maddox, S. J., Efstathiou, G., Sutherland, W. J., and Loveday, J. 1990, \cr
&{\it M.N.R.A.S} {\bf 242}, 43 \cr
\noalign{\vskip10pt}

&Meinhold, P. and Lubin, P. 1991, {\it Ap. J. Lett.} {\bf 370}, L11 \cr
\noalign{\vskip10pt}

&Peebles, P. J. E. 1967, {\it Ap. J.} {\bf 153}, 1 \cr
\noalign{\vskip10pt}

&Peebles, P. J. E. 1980, {\it The Large Scale Structure of the Universe} \cr
&(Princeton University Press) \cr
\noalign{\vskip10pt}

&Peebles, P. J. E. 1987, {\it Ap. J.} {\bf 315}, L73 \cr
\noalign{\vskip10pt}

&Reynolds, R. J. et al. 1986, {\it Astrophys. J. Lett.} {\bf 309}, L9 \cr
\noalign{\vskip10pt}

&Sachs, R. K. and Wolfe, A. M. 1967, {\it Ap. J.} {\bf 147}, 73 \cr
\noalign{\vskip10pt}

&Salati, P., Wallet, J. C. 1984, {\it Phys. Lett} {\bf B144}, 61 \cr
\noalign{\vskip10pt}

&Sciama, D. W., 1990a, {\it Ap. J.} {\bf 364}, 549 \cr
\noalign{\vskip10pt}

&Sciama, D. W., 1990b, {\it M.N.R.A.S.} {\bf 244}, 1 \cr
\noalign{\vskip10pt}

&Sciama, D. W., 1990c, {\it M.N.R.A.S.} {\bf 246}, 191 \cr
\noalign{\vskip10pt}

&Sciama, D. W., 1990d, {\it Nat.} {\bf 346}, 40 \cr
\noalign{\vskip10pt}

&Scott, D., Rees, M. J. and Sciama, D. W. 1991, \cr
&{Astron. Astrophys.} {\bf 250}, 295  \cr
\noalign{\vskip10pt}

&Shapiro 1986a, referenced in Giroux 1993 \cr
\noalign{\vskip10pt}

&Shapiro 1986b, referenced in Giroux 1993 \cr
\noalign{\vskip10pt}

&Shapiro and Giroux 1987, referenced in Giroux 1993 \cr
\noalign{\vskip10pt}

&Smooth, G. et al. 1992, {\it Ap. J. Lett} {\bf 396}, L1 \cr
\noalign{\vskip10pt}

&Songaila, A. et al. 1989, {\it Astrophys. J. Lett.} {bf 345}, L71 \cr
\noalign{\vskip10pt}

&Timbie, P. T. and Wilkinson, D. T. 1990, {\it Ap. J.} {\bf 353}, 140 \cr
\noalign{\vskip10pt}

&Tuluie, R. 1993a, PhD. Thesis, University of Texas \cr
\noalign{\vskip10pt}

&Tuluie, R. et al. 1993b, Work in progress \cr
\noalign{\vskip10pt}

&Vishniac, E. T. 1987, {\it Ap. J.} {\bf 322}, 597 \cr
\noalign{\vskip10pt}

&Vishniac, E. T. 1993, private communication \cr
\noalign{\vskip10pt}

&Weinberg, S. 1972, {\it Gravitation and Cosmology} \cr
&(Wiley and Sons, New York) \cr
\noalign{\vskip10pt}

&Zel'dovich, Ya. B. 1970, {\it Astron. Asrophys.} {\bf 5}, 84 \cr
\noalign{\vskip10pt}

&Zel'dovich, YA. B., {\it Relativistic Astrophysics},  Chicago Press 1971 \cr
\noalign{\vskip10pt}

&Zel'dovich, Ya. B. 1972, {\it M. N. R. A. S.} {\bf 160}, 1 \cr }

\end{document}